# Generalized Synchronization in Ginzburg–Landau Equations with Local Coupling

P. V. Popov, A. A. Koronovskiĭ, and A. E. Hramov*

*Saratov State University, Saratov, Russia*
* *e-mail: aeh@cas.ssu.runnet.ru*

**Abstract**—The establishment of generalized chaotic synchronization in Ginzburg–Landau equations unidirectionally coupled at discrete points of space (local coupling) has been studied. It is shown that generalized synchronization regimes are also established with this type of coupling, but the necessary intensity of coupling is significantly higher than that in the case of a spatially homogeneous coupling.

Synchronization of chaotic oscillations in dynamical systems is among important basic nonlinear phenomena that have been extensively studied in recent years [1]. Such investigations are both of considerable theoretical importance for deeper insight into the general laws of interaction between complex nonlinear systems of various natures (physical, chemical, biological, etc.) and of practical significance, for example, in the context of solving problems related to data transmission using deterministic chaos, analysis of neuron ensembles, diagnoses of disorders, etc. [2–4].

One important type of synchronous behavior that has received much attention is the generalized synchronization (GS) of unidirectionally coupled chaotic systems [5]. This type of synchronization implies that the state vectors of the drive ($\mathbf{x}_d(t)$) and response ($\mathbf{x}_r(t)$) chaotic systems upon termination of the transient process obey a certain functional relationship: $\mathbf{x}_r(t) = \mathbf{F}[\mathbf{x}_d(t)]$. The type of this function can be different (in particular, fractal [6]), and in most cases its explicit form remains unknown. Several approaches to the diagnostics of GS between unidirectionally coupled chaotic systems were described in the literature [5–7]. The GS regimes were previously studied in sufficient detail for systems with small numbers of degrees of freedom [5–10].

It is of considerable interest to study the GS phenomenon in spatially distributed chaotic systems [11–13]. Previously, we have analyzed the process of GS establishment in unidirectionally coupled Ginzburg–Landau equations with a spatially homogeneous diffusion coupling [12, 13]. However, it is also very important to study the synchronization and control of chaotic oscillations in coupled distributed systems in the case when an external signal acts upon the drive system at a certain finite number of points in space (local or point coupling) [14, 15]. This type of local spatial coupling was studied in detail in the case of complete chaotic synchronization of two identical distributed systems [14, 16].

The aim of this study was to analyze the possibility of establishing a GS regime in spatially distributed systems described by the Ginzburg–Landau equations with a local coupling introduced at a finite number of points in the space.

The mathematical model represents a system of two unidirectionally coupled one-dimensional complex Ginzburg–Landau equations,

$$\partial u/\partial t = u - (1 - i\alpha_d)|u|^2 u + (1 + i\beta_d)\partial^2 u/\partial x^2, \quad x \in [0, L], \quad (1)$$

$$\partial v/\partial t = v - (1 - i\alpha_r)|v|^2 v + (1 + i\beta_r)\partial^2 v/\partial x^2 + \varepsilon \mathcal{F}|u, v|, \quad x \in [0, L] \quad (2)$$

with the periodic boundary conditions $u(x, t) = u(x + L, t)$ and $v(x, t) = v(x + L, t)$, where $L$ is the spatial period and $i = \sqrt{-1}$. Equation (1) describes the distributed drive system, Eq. (2) refers to the distributed response system, $\varepsilon$ is a parameter characterizing the intensity of diffusion coupling, and $u, v$ is a function describing the unidirectional coupling between the drive and response systems. For this study, the latter function was selected in the following form:

$$\mathcal{F}[u, v] = \delta(x - N\Delta X)(u - v), \quad N = 0, 1, 2, \ldots, \quad (3)$$

where $\delta(\xi)$ is the Dirac delta function and $\Delta X$ is the distance between spatial points at which the systems are coupled (note that $\Delta X = 0$ corresponds to the case of a spatially homogeneous dissipative spatial coupling studied previously [12, 13]).

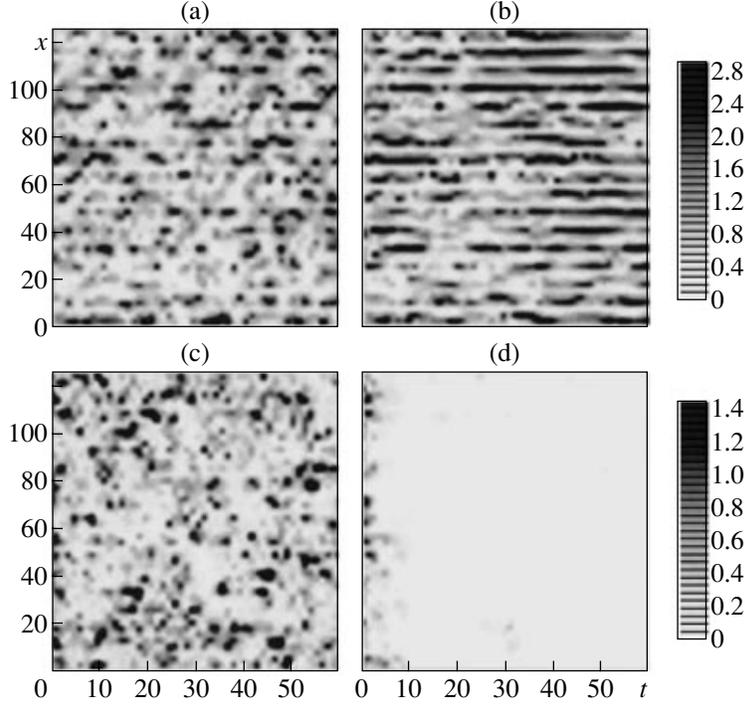

**Fig. 1.** Spatiotemporal diagrams of the amplitudes of the differences of states of (a, b) the drive and response systems, $|u(x, t) - v(x, t)|^2$ and (c, d) the response and auxiliary systems, $|v(x, t) - v_a(x, t)|^2$, for two values of the coupling parameter: (a, c) $\varepsilon = 20$, asynchronous regime; (b, d) $\varepsilon = 100$, GS regime.

The numerical modeling of Eqs. (1) and (2) was carried using an explicit computational scheme with $\Delta t = 0.0002$ and $\Delta x = L/1024$. Prior to switching on the coupling, the two subsystems remained uncoupled for a certain period of time. For this study, the parameters of drive ($\alpha_d$, $\beta_d$) and response ($\alpha_r$, $\beta_r$) systems under consideration were selected as follows: $\alpha_d = 1.5$, $\beta_d = 1.5$; $\alpha_r = 4.0$, $\beta_r = 4.0$; and the spatial period was set as $L = 40\pi$. It is known that autonomous distributed systems with such control parameters occur in the state of spatiotemporal chaos [12, 17].

The diagnostics of GS between the distributed drive and response systems was based on the auxiliary system method [7]. According to this method, an additional response system $v_a(x, t)$ identical to the original response system $v(x, t)$ described by Eq. (2) is introduced; its initial conditions $v_a(x, t_0)$ are different from $v(x, t_0)$. If GS is absent, vectors of the response $v(x, t)$ and auxiliary $v_a(x, t)$ systems are always different. In contrast, when GS takes place, the relations $v(x, t) = \mathbf{F}[u(x, t)]$ and $v_a(x, t) = \mathbf{F}[u(x, t)]$ are valid and, hence, the states of the response and auxiliary systems upon termination of the transient process must be identical: $v(x, t) \equiv v_a(x, t)$. Thus, the identity of states of the response and auxiliary systems upon termination of the transient process is a criterion for the existence of GS between the drive and response chaotic systems.

Figure 1 shows spatiotemporal distributions of the difference of states of the response and drive systems, $|v(x, t) - u(x, t)|^2$, and the response and auxiliary systems, $|v(x, t) - v_a(x, t)|^2$, for a distance between the local coupling points $\Delta X = 7.36$ and different coupling parameters $\varepsilon = 20$ and 100. As can be seen, oscillations of the response and drive systems are different in both cases and their difference does not reveal the presence of synchronous dynamics. A different situation is observed for the comparative analysis of the behavior of the response and auxiliary systems. Indeed, for a smaller value of the coupling parameter $\varepsilon$, the states of the response and auxiliary systems are different (Fig. 1c), which is indicative of the absence of a GS regime. As the parameter of coupling between the distributed systems with unidirectional local coupling is increased, GS is established as illustrated by Fig. 1d, which clearly indicates that the systems occur in identical spatiotemporal states $v(x, t) \equiv v_a(x, t)$ after a short transient process.

Previously, it was established [12, 13] that a threshold value of the coupling parameter ($\varepsilon_{GS}$) for the appearance of GS in the case of a spatially homogeneous coupling weakly depends on the control parameters of the drive system. Analogous results were obtained for the local coupling between systems under consideration. On the other hand, analysis showed that the intensity of coupling at which the GS appears is significantly dependent on the degree of inhomogeneity of this coupling (i.e., on the distance $\Delta X$ between the points of coupling). This is illustrated in Fig. 2a, which shows a plot of the coupling threshold $\varepsilon_{GS}$ for GS onset versus coupling inhomogeneity parameter $\Delta X$. As can be seen, an increase in the inhomogeneity of coupling (i.e., in the distance $\Delta X$ between the points of coupling) is accompanied by a rapid increase in the coupling threshold $\varepsilon_{GS}$ (at $\Delta X = 0$, GS appears at $\varepsilon_{GS} \approx 0.76$).

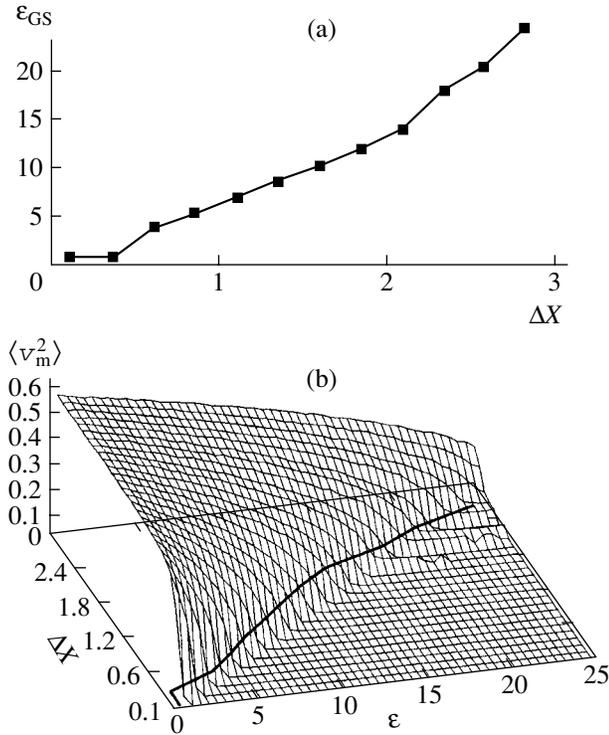

**Fig. 2.** diagrams of (a) the coupling threshold $\varepsilon_{GS}$ for the GS onset versus coupling inhomogeneity parameter $\Delta X$ and (b) the average square amplitude (power) of oscillations of the modified response system $v_m$ versus the coupling intensity ($\varepsilon$) and inhomogeneity ($\Delta X$) parameters. Thick solid line indicates the boundary of the GS onset (analogous to the curve in Fig. 2a).

It was demonstrated [12] that the establishment of GS in coupled Ginzburg–Landau equations can be conveniently traced by analyzing a modified system which differs from the initial one by an additional dissipative term in the response subsystem. In the case of a spatially inhomogeneous local diffusion coupling, such a modified system can be written as follows:

$$\partial v_m/\partial t = v_m - (1 - i\alpha_r)|v_m|^2 v_m + (1 + i\beta_r)\partial^2 v_m/\partial x^2 + \varepsilon \mathcal{F}[0, v_m], \quad x \in [0, L]. \quad (4)$$

According to the previous analysis [9, 12], a GS regime in coupled dynamical systems appears when (i) the modified system passes from chaotic to a stationary (or periodic) regime and (ii) the amplitude of the external action strongly exceeds the amplitude of self-oscillations $v_m(x, t)$ in the modified system, so that the spatiotemporal state of the modified response system is shifted in the phase space toward regions corresponding to strong dissipation.

Figure 2b shows a plot of the average square amplitude (power) $\langle v_m^2 \rangle$ of oscillations in the modified Ginzburg–Landau system as a function of the coupling parameter $\varepsilon$ and the inhomogeneity parameter $\Delta X$. Similarly to the case of spatially homogeneous coupling (see [12]), GS appears when the modified response system exhibits chaotic oscillations whose amplitudes at which a fixed $\Delta X$ value decreases with increasing $\varepsilon$ (Fig. 2b). Simultaneously, the external signal power (proportional to $\varepsilon^2$) begins to significantly exceed the power $\langle v_m^2 \rangle$ of intrinsic oscillations of the modified response system. As the inhomogeneity parameter $\Delta X$ increases (i.e., the number of coupling points decreases), the additional dissipation introduced into the response system drops and, accordingly, the threshold coupling parameter $\varepsilon_{GS}$ grows (Fig. 2b). However, the average power $\langle v_m^2 \rangle$ corresponding to GS onset (i.e., to the threshold coupling intensity $\varepsilon_{GS}$) remains approximately the same (thick solid line in Fig. 2b) in a broad range of variation of the inhomogeneity parameter $\Delta X$. This result indicates that the GS threshold $\varepsilon_{GS}$ is determined predominantly by the dynamics of the modified distributed system (4) with additional dissipation.

To summarize, we have studied the establishment of a GS regime in a system of unidirectionally coupled distributed autooscillatory media described by complex Ginzburg–Landau equations with local spatial coupling. It has been established that the onset of GS in this case is determined generally by the same mechanisms as those in the system with spatially homogeneous coupling studied previously. On the other hand, the threshold of the onset of GS strongly depends on the degree of inhomogeneity of the local coupling.

**Acknowledgments.** This study was supported in part by the Russian Foundation for Basic Research (project nos. 05-02–16273 and 06-02-16451) and the "Dynasty" Foundation for Noncommercial Programs.